# Certified to Drive: A Policy Proposal for Mandatory Training on Semi-Automated Vehicles

Soumita Mukherjee

Pennsylvania State University, szm6423@psu.edu

Varun Darshana Parekh

Pennsylvania State University, vdp5074@psu.edu

Nikhil Tayal

Tezmee Inc. nikhil@tezmee.com

**Executive Summary**

According to the Society of Automotive Engineers (SAE) J3016 standard, Advanced Driver Assistance Systems (ADAS) refer to a suite of technologies that automate specific driving tasks (such as acceleration, braking, and steering) while still requiring different levels of active driver oversight. While nearly 92.7% of new U.S. vehicles now include at least one automated driving feature, the transformative potential lies in the integration of advanced systems—specifically, ADAS Level 2. While these systems promise enhanced safety and efficiency, they have also created a significant knowledge gap. Drivers often lack the necessary training to understand system limitations and intervene effectively, leading to accidents and liability issues that undermine the benefits of automation. This paper highlights the urgent need for policies to address this gap. Although the Boeing 737 Max incidents resulted from a mix of design shortcomings, regulatory oversights, and systemic issues, they also highlight a critical gap in pilot training on managing automated systems during abnormal conditions. This example demonstrates the urgent need for focused, concise training on human-automation interaction—a need that is equally critical for operators of Level 2 ADAS-equipped vehicles, as discussed in detail later in this article. The lack of structured education for semi-automated vehicle (SAV) operators mirrors similar risks in other industries, where formal training is critical for safe operation. Two policy recommendations are proposed. First, governments should create concise, official resources in accessible and official format to educate drivers on system capabilities and limitations. Second, mandatory training and certification programs should be introduced, combining theoretical and hands-on components to prepare drivers for real-world scenarios. These measures will improve driver understanding, reduce misuse, and foster public trust in semi-automated vehicle technologies. By addressing the knowledge gap, policymakers can ensure a safer, more responsible transition to automation, maximizing its benefits while minimizing risks to public safety.

**INTRODUCTION**

ADAS (Advanced Driver-Assistance Systems) has a formal, well-defined meaning - , the practical examples of semi-automated cars illustrate the true "human in the loop" aspect. Tesla's Autopilot, Ford's BlueCruise, GM's Super Cruise, BMW's Personal Pilot L3 and Highway Assistant, Jeep's Hands-Free Active Driving Assist, Nissan's ProPilot Assist 2, and Toyota's Teammate all demonstrate how a vehicle can take over some driving tasks—often allowing for hands-free operation—yet still require the driver's active oversight and readiness to intervene. In other words, these systems blend automation with human responsibility, underscoring the critical role of the driver even when advanced technology is at the wheel. The adoption of Advanced Driver Assistance Systems (ADAS) in the United States has surged, with 92.7% of new vehicles now equipped with at least one such feature (AAA, 2019). As defined in Society of Automotive Engineers (SAE) J3016 standard, ADAS has 6 different levels of autonomy ranging from extremely driver active, no autonomy level 0 to completely autonomous level 6 which are yet to be commercialised. This widespread integration reflects a pivotal shift in driving dynamics, as traditional responsibilities are increasingly transferred to automated systems. Drivers, once the primary agents of control, are now positioned as supervisors tasked with monitoring these systems and intervening when necessary. However, this transition from active control to passive oversight necessitates a baseline understanding of how these systems operate. Drivers' familiarity is shaped through exposure to instructional materials, training sessions, hands-on use, and cumulative experience. Yet, as the adoption of partially automated systems accelerates, a critical question emerges: Why are drivers disproportionately blamed when automation fails? Challenges such as overtrust, misjudgment of system capabilities, or misunderstanding its limitations are often attributed to users. However, these issues are less about individual fault and more about systemic shortcomings in education and training. Addressing this gap is crucial to fostering equitable accountability and ensuring the safe integration of automation into daily transportation.

Historical technological disasters—such as the Boeing 737 MAX crashes, the Three Mile Island nuclear incident, the Bhopal gas tragedy, and the Chernobyl disaster—illustrate that, despite differences in nature, environment, and triggering events, a common thread has been the failure to adequately train human operators. The literature (Meshkati, 1991; Malone, 1990; Spielman & Blanc, 2020) emphasizes that beyond human negligence it is the inefficient training that were pivotal factors in these tragedies, ultimately driving significant reforms in operator training mandates across various industries. These diverse cases convincingly underscore that it is not merely operator negligence but rather insufficient training and poorly designed systems that can precipitate catastrophic outcomes (Meshkati, 1991). The integration of automation, particularly in fields like aviation and automotive systems, amplifies the importance of addressing human factors to prevent accidents. As automation increasingly takes on critical roles, ensuring robust training and designing systems and having policies in place that account for human interaction become indispensable to preventing future tragedies.

Similar challenges arise in semi-autonomous vehicles, where drivers struggle to transition from operators to supervisors (Mersinger & Chaparro, 2022). Research emphasizes the need for human-centered design, improved communication between automated systems and operators, and comprehensive training programs (Blackett, 2021; Strauch, 2017; Sheridan & Nadler, 2006). To ensure safety,establishing minimum driver training standards for partially automated cars, drawing lessons from the aviation industry's successful integration of automation (Casner & Hutchins, 2019) is crucial.

This policy proposal is intended for presentation to state governments and Departments of Transportation (DOTs). The aim is to advocate for their leadership in implementing regulations and standards for semi-automated vehicle training and driver certification. By aligning this initiative with transportation safety goals, it addresses the growing risks associated with semi-automated vehicle usage and promotes a framework for safer adoption of this transformative technology.

Therefore, it is imperative to consider the broader implications of automation and its interaction with human operators. Policies must address existing gaps by enhancing operator knowledge and fostering effective human-automation collaboration. In this context, a strategic policy overhaul could bridge the divide between drivers and partially automated systems.

The policy options outlined address the following issues:

- A need for governments to regulate the safe use of semi-automated vehicles.
- Ensuring Level 2 automation works within its limitations with driver involvement.
- Requiring driver training to prevent over trust and misuse of automation.

**POLICY OPTIONS**

Policy Option 1: Creating Official, Byte-Sized Information Resources

**Description:**

To bridge the knowledge gap among operators of semi-autonomous vehicles, developing official, easily accessible educational resources is essential. The resources should be directed from the individual state DOTs in the form of additional training and tests being included in the existing driving courses for individuals getting licensed to drive. Traditional methods like reading owner's manuals or simply observing vehicle features are often ineffective for adults learning new driving technologies, especially level 2 ADAS systems. Research indicates that drivers who engage with their preferred learning methods—particularly hands-on demonstrations—exhibit higher understanding and increased usage of in-vehicle systems (Hillary, 2018). A study showed that observing Advanced Driver Assistance Systems (ADAS) during demonstration drives significantly enhances positive perceptions, knowledge, and trust in the technology compared to just reading about it (Nylen, 2019). Therefore, aligning learning approaches with driver preferences and incorporating interactive, practical experiences are crucial for effectively teaching adults about the Level 2 ADAS systems.(L. Fridman et al., 2019).

Currently, operators often rely on unofficial sources such as YouTube, Reddit, or social media, which can vary widely in accuracy and may not convey critical safety information or limitations of the technology. Creating "byte-sized" information resources—brief, clear, and easily digestible— in the form of official short documents and videos would enable drivers to understand essential aspects of semi-autonomous systems without feeling overwhelmed.

States like California, Virginia, and Georgia offer a comprehensive series of educational and instructional videos for drivers. However, this currently doesnot include information on semi-autonomous training. Hence inclusion of informational videos for semi-automated vehicles should focus on the systems' limitations and provide practical guidance on handling common situations, ensuring the information is more accessible and is also engaging than traditional, dense manuals.

These materials would be most effective if grounded in principles of socio-centric technical writing, bridging the gap between complex technical details and user-friendly language. This approach ensures that both the technology and the social context in which it operates are considered, making the information relatable to a diverse audience of drivers. Including interactive elements, such as quizzes or quick assessments, can ensure drivers actually engage with the material rather than merely skimming through it. Relevant examples could include short videos that demonstrate common semi-autonomous scenarios and guide drivers on appropriate responses.

**Advantages:**

- Increased Accessibility: Byte-sized resources are easy to access and consume, making it more likely that drivers will engage with the material.
- Enhanced Understanding: Concise and clear information improves comprehension of complex technologies.
- Cost-Effectiveness: Developing digital resources is generally less expensive than organizing extensive training programs.
- Wide Reach: Online materials can be distributed broadly, reaching a larger audience without geographical limitations.
- Flexibility: Drivers can learn at their own pace and revisit materials as needed.

**Challenges:**

- Variable Engagement: Without mandatory requirements, some drivers may choose not to utilize these resources.
- Information Overload: Even concise materials may be overlooked if drivers feel inundated with content.
- Quality Assurance: Ensuring the accuracy and reliability of the information is crucial to avoid misinformation.
- Digital Divide: Not all drivers may have access to digital platforms or possess the digital literacy required to engage with online materials.
- Limited Interactivity: While informative, byte-sized resources may not fully replicate the benefits of hands-on demonstrations and practical experience.

**Supporting Actions:**

- Mandate periodic updates to these resources to reflect the latest SAV developments and safety protocols.
- Promote engagement through interactive elements (like quizzes) that enhance understanding and ensure drivers retain the information.
- Publicize benefits by partnering with insurance companies to offer discounts for drivers who complete these informational modules.
- Evidence from similar informational programs, like those in the aviation or commercial driving sectors, could support this recommendation. Additionally, studies on the efficacy of microlearning modules and chunked learning for adult learners would provide evidence that these materials can improve driver understanding and response times.

Policy Option 2: Establishing Training Programs and Certification Requirements

**Description:**

Given the limited user understanding of Level 2 autonomous vehicles—including how the system operates, its interaction points with the driver, and the boundaries of its capabilities and limitations—there is a clear need for targeted training to ensure safe and effective use of semi-automated driving technologies. Implementing mandatory training and certification programs for operators of semi-autonomous vehicles is essential to enhance driver competence and road safety. A structured, hands-on training approach would allow drivers to experience firsthand how semi-autonomous systems operate, understand their limitations, and recognize situations that may require human intervention. These programs could be modeled after existing training requirements for commercial vehicle operators, such as the "Minimum Training Requirements for Entry-Level Commercial Motor Vehicle Operators" (49 CFR Part 380, March 18, 2022).

An illustrative example is the widespread use of defensive driving courses, which aim to improve road safety by equipping drivers with advanced skills to anticipate and respond to potential hazards. Many states have enacted laws encouraging or requiring insurers to offer discounts upon completion of these courses. Specifically, 37 out of 50 states mandate auto insurance discounts for drivers who complete approved defensive driving programs. These initiatives incentivize safer driving behaviors through enhanced driver education. Another example is New York that requires new drivers to complete a pre-licensing course before the driving test.

Similarly, certification courses tailored for semi-autonomous driving technology could be established. This approach would mitigate the risks associated with relying solely on in-car manuals or informal online resources, ensuring that drivers are adequately prepared to operate advanced automation systems safely.

**Advantages:**

- Improved Driver Competence: Formal training ensures drivers have a comprehensive understanding of semi-autonomous systems, their operational capabilities, and limitations.
- Improved Road Safety: Educated drivers are better equipped to handle unexpected situations, reducing the likelihood of accidents due to misuse or misunderstanding of automation features.
- Standardization of Knowledge: Certification creates a baseline level of expertise among all operators of semi-autonomous vehicles, promoting consistency in how these vehicles are used.
- Incentivization through Insurance Discounts: Similar to defensive driving programs, insurers could offer premium reductions for certified drivers, encouraging broader participation.
- Facilitation of Legal and Regulatory Compliance: Trained drivers are more likely to adhere to laws and regulations governing the use of semi-autonomous vehicles.

**Challenges:**

- Implementation Costs: Developing, administering, and maintaining training programs and certification processes require substantial financial and logistical resources at the State level.
- Accessibility Issues: Ensuring equitable access to training programs for all drivers, including those in rural or underserved areas or other vulnerable population, may be challenging.
- Resistance to Mandatory Requirements: Some drivers may oppose compulsory training and certification, viewing it as an unnecessary burden or infringement on personal freedoms.
- Rapid Technological Evolution: Semi-autonomous vehicle technology is continually advancing, necessitating frequent updates to training curricula to remain current.
- Enforcement Complexity: Monitoring compliance and enforcing certification requirements would demand robust administrative systems and could strain existing regulatory bodies.

**Supporting Actions:**

- Collaborate with SAV manufacturers to develop a consistent curriculum that aligns with system specifications across different models.
- Offer incentives, such as insurance premium reductions, for certified drivers, and require certification as a condition for SAV insurance coverage.
- Require re-certification every few years or upon significant SAV updates to ensure drivers remain informed of the latest system capabilities and limitations.
- Similar training and certification requirements in fields where individuals operate complex systems (e.g., piloting, commercial trucking) show that well-trained operators have fewer incidents. Citing statistics on reduced accident rates in certified drivers, as well as evidence from sectors where mandatory training exists, could underscore the need for this policy.

**CONCLUSION**

As semi-automated vehicle technologies continue to revolutionize transportation, the need for robust policies ensuring safe and informed adoption becomes increasingly urgent. This paper highlights the critical gap in driver knowledge and training, which poses significant risks to the safety and effectiveness of these systems. By implementing solutions such as concise informational resources, mandatory training programs, and state-level regulatory oversight, governments and transportation agencies can address these challenges comprehensively.

These measures not only improve driver understanding and engagement with automation but also foster public trust in the technology. Establishing standards for training and certification ensures accountability and reduces the potential for misuse, paving the way for a safer transition to a future where automation plays a central role in mobility. Collaboration among policymakers, manufacturers, and educational institutions will be essential in creating a framework that prioritizes safety and sets a precedent for responsibly integrating emerging technologies into society.

# References

Meshkati, Najmedin. 'Human Factors in Large-Scale Technological Systems' Accidents: Three Mile Island, Bhopal, Chernobyl'. Industrial Crisis Quarterly 5, no. 2 (1991/6): 133–54. https://doi.org/10.1177/108602669100500203.

Noah, T. Curran, W. Kennings Thomas, and Shin K. 'Analysis and Prevention of MCAS-Induced Crashes', 2023.

Spielman, Zachary, and Katya Le Blanc. 'Boeing 737 MAX: Expectation of Human Capability in Highly Automated Sy Stems'. In Advances in Intelligent Systems and Computing, 64–70. Springer International Publishing, 2020. https://doi.org/10.1007/978-3-030-51758-8_9.

C Mersinger, Molly, and Alex Chaparro. Semi-Autonomous Vehicle Crashes: An Exploration of Contributing Factor s. AHFE International, n.d. https://doi.org/10.54941/ahfe1002471.

Strauch, Barry. 'The Automation-by-Expertise-by-Training Interaction'. Human Factors: The Journal of the Human Factors and Ergonomics Society 59, no. 2 (27 September 2016): 204–28. https://doi.org/10.1177/0018720816665459

T., Sheridan, and Nadler E. 'A Review of Human-Automation Interaction and Lessons Learned', 2006.

Blackett, Claire. 'Human-Centered Design in an Automated World'. In Advances in Intelligent Systems and Computing, 17–23. Springer International Publishing, 2021. https://doi.org/10.1007/978-3-030-68017-6_3.

Casner, Stephen M., and Edwin L. Hutchins. 'What Do We Tell the Drivers? Toward Minimum Driver Training Standards for Partially Automated Cars'. Journal of Cognitive Engineering and Decision Making 13, no. 2 (8 March 2019): 55–66. https://doi.org/10.1177/1555343419830901.

Nylen, Ashley B., Michelle L. Reyes, Cheryl A. Roe and Daniel V. McGehee. "Impacts on Driver Perceptions in Initial Exposure to ADAS Technologies." Transportation Research Record 2673 (2019): 354 - 360.

Fridman, Lex, Daniel E. Brown, Michael Glazer, William Angell, Spencer Dodd, Benedikt Jenik, Jack Terwilliger, et al. 'MIT Advanced Vehicle Technology Study: Large-Scale Naturalistic Driving Study of Driver Behavior and Interaction With Automation'. IEEE Access 7 (2019) 102021–38. https://doi.org/10.1109/ACCESS.2019.2926040

Abraham, Hillary, B. Reimer and Bruce Mehler. "Learning to Use In-Vehicle Technologies: Consumer Preferences and Effects on Understanding." Proceedings of the Human Factors and Ergonomics Society Annual Meeting 62 (2018): 1589 - 1593.

American Automobile Association. 2019. *Advanced Driver Assistance Technology Names*. https://www.aaa.com/AAA/common/AAR/files/ADAS-Technology-Names-Research-Report.pdf

SAE International. (2023). *Active safety lane keep assist system (automated lateral vehicle motion support) – Passenger car, MPV, and light truck (SAE Standard No. J3262_202312)*